\documentclass{ws-ijmpa}

\begin{document}

\markboth{P.D.Stack and R.Delbourgo}
{GR of colour-property}
\catchline{}{}{}{}{}

\title{The Relativity of Colour}

\author{Paul D Stack and Robert Delbourgo}

\address{School of Physical Sciences, University of Tasmania, 
Locked Bag 37 GPO\\ Hobart, Tasmania 7001,AUSTRALIA 7001\\
pdstack@utas.edu.au, bob.delbourgo@utas.edu.au}

\maketitle
 
\begin{abstract}
By attaching three anticommuting Lorentz scalar (colour) property coordinates to space-time, 
with an appropriate extended metric, we unify gravity with chromodynamics: 
gauge transformations then just correspond to coordinate transformations in the 
enlarged spacetime-property space.
\end{abstract}
\maketitle

\section{\label{intro} Introduction}
We have come a long way since the work of Kaluza and Klein on the unification of gravity
with electromagnetism obtained by attaching a fifth coordinate to the four spacetime coordinates. 
By enlarging the spacetime with further bosonic coordinates which, like the fifth, 
are `curled up' it is possible to attain a unified description of gravity with Yang-Mills theory,
and the strong interactions in particular --- but at the price of producing infinite towers of
excitations of the ground state\cite{C,SS,DL,CK,OW}. One may even extend the idea with
supersymmetry\cite{DNP} but at the extra price of some spin state proliferation. 

In recent years we have suggested an alternative enlargement based on Lorentz scalar 
anti-commuting coordinates, which avoids towers of excitations and spin problems.
We have associated these new coordinates with concrete `attributes' that accompany
events. The next step in our endeavour to unify gravity with the other forces involves the consideration of the strong force. As in our previous papers\cite{paper1,paper2,paper3}, 
we achieve this through the addition of property coordinates to space-time in order to 
describe the ``what'' that is missing from the ``when'' and ``where'' of an event. 
In the case of the strong force we need to include three `chromicity' coordinates, 
one for each colour.  Since the strong force conserves parity we do not need to consider
 separate left and right chiral property coordinates, simplifying the situation greatly.  

So far our work on this topic has been as follows: first\cite{paper1} we carefully worked through how 
to extend general relativity with anti-commuting property coordinates and then applied it to a 
one coordinate model, or electricity. This unified gravity with electromagnetism and also 
produced a cosmological constant, the sign of the which was wrong due to a lack of freedom in 
our metric -- but this was only an issue for the one coordinate case. Next\cite{paper2} we 
considered two property coordinates and found that we could unify gravity with a U(2) 
Yang-Mills gauge field, again producing a cosmological constant but this time with enough 
freedom that it was at least consistent with observation. To properly treat the weak force one 
needs to consider chiral fields, so finally\cite{paper3} we looked again at two property coordinates, 
but assigned them left and right chiralities. This resulted in unification of gravity with 
U(1)${}_L\times$U(1)${}_R$ gauge fields and a cosmological constant. To ensure the 
universality of gravitational interactions we found the gauge couplings had to be equal at 
high energy.

The next step, which is the focus of this paper, is to consider three property coordinates associated
with `chromicity' or colour and to attempt to unify gravity with the strong force. In Section \ref{sec2}
we will briefly introduce property coordinates, then Section \ref{sec3} covers the construction of a supermetric and the resulting Lagrangian that is produced. In Section \ref{sec4} we will describe 
how matter fields come into play and finally in Section \ref{sec5} we offer some conclusions and 
mention what steps to take next.

\section{\label{sec2} Colour property}
We attach three complex anti-commuting coordinates $\zeta$ and their conjugates $\bar\zeta$ 
to space-time. We choose these coordinates to be Lorentz scalars, which distinguishes our 
scheme from supersymmetry where the extra degrees of freedom transform as spinors, leading 
to spin state proliferation.

As we have done previously, we adopt the convention that space-time-property indices run over uppercase Roman characters ($M$, $N$, $L$, etc), space-time indices run over lower case 
Roman characters ($m$, $n$, $l$, etc) and property indices run over Greek letters ($\mu$, $\nu$, 
$\lambda$, etc). Thus our superspace is denoted by $X^M = (x^m, \zeta^\mu, \zeta^{\bar\mu})$.

In our last paper\cite{paper3} we discussed left and right handed chiralities, but since the strong 
force conserves parity we do not need that distinction here. We assign quantum numbers to 
our chromic coordinates as follows: $(\zeta^1, \zeta^2, \zeta^3)$, Colour = (Red, Green, Blue), 
Quark number = (1/3, 1/3, 1/3) and similarly for the conjugates: $(\zeta^{\bar 1}, \zeta^{\bar 2},
\zeta^{\bar 3})$, Colour = (Anti-Red, Anti-Green, Anti-Blue), Quark number = (-1/3, -1/3, -1/3). 
The matter fields resulting from this assignment are discussed later in Section \ref{sec4}.

\section{\label{sec3} Supermetric and Lagrangian}
The construction of the metric follows a very similar procedure to that of our U(2) 
paper\cite{paper2}, we start with a flat manifold possessing coordinates 
${X^M} = (x^m,\zeta^\mu,\zeta^{\bar\mu})$ without space-time curvature or gauge fields. 
Our metric distance in this case is given by:
\[ ds^2 = dX^M dX^N\eta_{NM} = dx^m dx^n\eta_{nm} +  \ell^2 (d\zeta^\mu 
d\zeta^{\bar\nu} \eta_{\bar\nu \mu}+d\zeta^{\bar\mu} d\zeta^\nu \eta_{\nu\bar\mu})/2
 \]
where $\eta_{\mu \bar\nu} = -\eta_{\bar\nu \mu} = \delta_\mu{}^\nu $ and $\eta_{nm}$ is 
Minkowskian. The property coordinates $\zeta$ and $\bar\zeta$ are being taken as dimensionless, 
so a length scale $\ell$ has been introduced to ensure the metric distance carries the correct 
units.

We now deviate slightly from our previous work and consider the addition of the property 
curvature before the inclusion of gauge fields. The idea is to include U(3) invariants multiplied by 
free parameters $c_i$; these have no effect on the {\em global} U(3) invariance of
the metric distance.  In the U(2) paper\cite{paper2} we had 4 free parameters, in the 
U(1)$\times$U(1) paper\cite{paper3} we had 9. If we continue this to the U(3) case we would 
have 6 free parameters. However all we need is the freedom to allow Newton's constant and 
the cosmological constant to take sensible values. The restriction of these extra parameters
then becomes necessary as we move forward; we are still investigating possible ways of doing 
this but for now we will include one overall factor $C = 1 + c_1 \bar\zeta \zeta + c_2 (\bar\zeta \zeta)^2 + c_3 (\bar\zeta \zeta)^3 $. Thus our metric becomes:

\begin{equation}
\eta_{MN} = 
\left(
\begin{array}{ccc}
C \eta{}_{m}{}_{n}  & 0 & 0 \\ 
0 & 0 & \frac{1}{2} C  l^2 \delta{}^{\nu}{}^{\bar\mu} \\ 
0 & -\frac{1}{2} C l^2 \delta{}^{\mu}{}^{\bar\nu} & 0
\end{array}
\right)
\end{equation}

We then apply the formula from our first paper\cite{paper1} to get the Christoffel symbols:
\begin{equation}
2 \Gamma_{MN}{}^K \!\!=\!\!\big[ (-1)^{[M][N]} G_{ML,N}
  +  G_{NL,M}- (-1)^{[L]([M]\!+\![N])} G_{MN,L} \big](-1)^{[L]} G^{LK}.
\end{equation}
This is then used in the Palatini form of the Ricci scalar:
\begin{align}
{\cal R} = (-1)^{[L]}G^{MK}[(-1)^{[L][M]}{\Gamma_{KL}}^N{\Gamma_{NM}}^L
                - {\Gamma_{KM}}^N{\Gamma_{NL}}^L]
\end{align}
to produce the Lagrangian (via Mathematica):
\begin{equation}
\int \sqrt{G_{..}} \mathcal{R}\; d^3\zeta d^3\bar\zeta=-\frac{1152}{l^8}\sqrt{-g}K_\Lambda,
\end{equation}
where $K_ \Lambda= 5 c_1{}^4-10 c_1{}^2 c_2+2 c_2{}^2+3 c_1 c_3$. 

The inclusion of the property curvature factor $C$ has thus generated a cosmological constant. 
The next step is to introduce space-time curvature by generalizing from $\eta_{mn}$ to $g_{mn}$ 
and also to admit local U(3) phase transformations to our chromic coordinates so as to incorporate
QCD; these transformations of chromicity take the form:
\begin{equation} \label{propertyphasetrans}
x \rightarrow x;  \;\;\;
\zeta^{\mu} \rightarrow [e^{i \Theta(x)}]^{\mu\bar\nu} \zeta^\nu; \;\;\;
\zeta^{\bar\mu} \rightarrow \zeta^{\bar\nu} [e^{-i \Theta(x)}]^{\nu\bar\mu}.
\end{equation}
For our metric to transform correctly under such local phase transformations the minimal way is to introduce non-abelian gluon fields $W_{m}$, leading to the following metric:
\begin{equation}
G_{MN} \!=\! C
\left(
\begin{array}{ccc}
\! g{}_{m}{}_{n}\!+\!\frac{1}{2} g_w{}^2  l^2  \bar\zeta (W_{m}W_{n}\!+\!W_{n}W_{m}) \zeta  &
 -\frac{1}{2} i g_w l^2  (\bar\zeta W_{m}){}^{\bar{\nu}} & -\frac{1}{2} i g_w l^2  (W_{m}\zeta){}^{\nu} \\ 
-\frac{1}{2} i g_w l^2  (\bar\zeta W_{n}){}^{\bar{\mu}}& 0 & \frac{1}{2} l^2  \delta{}^{\nu}{}^{\bar\mu} \\ 
-\frac{1}{2} i g_w l^2   (W_{n}\zeta){}^{\mu} & -\frac{1}{2} l^2   \delta{}^{\mu}{}^{\bar\nu} & 0
\end{array}.
\right)
\end{equation}
There are two reasons for the choice of metric in the spacetime-property sector: we have to ensure
that it is overall fermionic and incorporates carriers of property, namely the gauge fields. One could
contemplate including in that sector ghost fields $V_m$ with the wrong spin-statistics multiplying
polynomials of $\bar{\zeta}\zeta$, but these have no place at the semiclassical level and it is hard
to see where they fit in at this stage; however they may have a role to play when quantizing gravity
via BRST\cite{F,DM,S}, but $V_m$ most certainly should {\em not} appear as asymptotic states;
similar remarks can be made for the scalar ghosts\cite{KO} helpful in quantizing gauge models.

Making use of Mathematica, the Lagrangian arising from (6) then comes out to be:
\begin{align}
\int \sqrt{G_{..}} R\; d^3\zeta d^3\bar\zeta =&\frac{4}{l^4}\sqrt{-g} \bigg( \frac{24}{l^2} K_gR^{[g]}
-\left(2c_2 - 3c_1{}^2\right) g_w{}^2 \mathcal{F}^{mn} \mathcal{F}_{mn}
 - \frac{288}{l^4}K_\Lambda \bigg)
\end{align}
where $\mathcal{F}_{mn} = W_{n,m} - W_{m,n} - i g_w [W_{m}, W_{n}]$ and 
$K_g=2c_1{}^3 -3c_1c_2+c_3$. This gives:
\begin{equation}
\kappa = g_w{}^2 l^2 (2c_2 - 3 c_1{}^2)/ 12 K_g, \quad \Lambda = 6 K_\Lambda / l^2 K_g.
\end{equation}
Even with this fairly minimal set of $c_i$ coefficients, there is easily enough freedom to fix 
the gravitational and cosmological constants to be what they physically are, though the 
smallness of $\Lambda$ implies fine-tuning.

\section{\label{sec4}Chromic polynomials and Matter Fields}
Focussing on the three chromic $\zeta$, we are dealing with an 
Sp(6) group that contains U(3)$\supset$ SU(3)$_{\rm colour}$. If
there are no powers of $\zeta$ or $\bar{\zeta}$, we are dealing 
with a black or colourless object, whereas the coefficient of
the cubic term $\zeta^1\zeta^2\zeta^3$ corresponds to a white 
system (or baryon); this is to be contrasted with the colour neutral
combination $\zeta^{\bar\sigma}\zeta^\sigma$ (a meson). However from the 
point of view of the colour group, we can make use of the Levi-Civita tensor 
$\epsilon^{\rho\sigma\tau}$ and its conjugate to contract/raise/lower
chromic indices; as far as SU(3) colour is concerned white is the 
same as black and all free hadrons have this property, but
chromicity regards them differently. Nonetheless the basic polynomials
$1, Z, Z^2/2!, Z^3/3!$ with $Z\equiv \zeta^{\bar{\sigma}}\zeta^\sigma$, which have 
already been used in introducing property curvature previously, are both colour and 
chromicity neutral and have the potential of describing `generations' of singlets.

\subsection{Colour and Generations}
To investigate this notion systematically, start with the linear in
$\zeta$ colour or chromic triplet $(\zeta^1,\zeta^2,\zeta^3)$
which belongs to the superfermion field $\Psi$.
Next consider the quadratic combination $(\zeta)^2$ -- part of a
bosonic superfield $\Phi$ -- corresponding to the colour product: 
$\bar{3}\times\bar{3} = 3 \oplus \bar{6}$, of which the $\bar{6}$ 
is missing because of anticommutativity!
Next add one power of $\bar{\zeta}$, leading to $3\times 3=\bar{3}
\oplus 6$. Here both representations are permitted; the new 
triplet can be written as $(\zeta^1,\zeta^2,\zeta^3)(\zeta^{\bar\sigma}
\zeta^\sigma)$ while the new sextet contains $\zeta^{\bar 3}
\zeta^1\zeta^2\sim \zeta^{\bar 3}\chi^{\bar 3}$, etc. We can envisage
another fermionic triplet through the polynomial $(\zeta)^3
(\bar{\zeta})^2$ having components
\begin{equation*}
\left(\zeta^1(\zeta^{\bar 2}\zeta^2\zeta^{\bar 3}\zeta^3),
      \zeta^2(\zeta^{\bar 3}\zeta^3\zeta^{\bar 1}\zeta^1),
      \zeta^3(\zeta^{\bar 1}\zeta^1\zeta^{\bar 2}\zeta^2)\right).
\end{equation*}
So far as bosons are concerned, in addition to the colour singlets 
and octets, contained in the combinations, $\zeta^{\bar\rho}\zeta^\sigma,
\zeta^{\bar\rho}\zeta^\sigma(\zeta^{\bar\tau}\zeta^\tau)$, there are only two
colour triplets:
\begin{equation*}
(\zeta^{\bar 2}\zeta^{\bar 3},\zeta^{\bar 3}\zeta^{\bar 1},
\zeta^{\bar 1}\zeta^{\bar 2})\quad{\rm and}\quad
(\zeta^{\bar 1},\zeta^{\bar 2},\zeta^{\bar 3})(\zeta^1\zeta^2\zeta^3),
\end{equation*}
plus their conjugates.

To summarise all these facts, it serves to draw up a table of 32
component fields each arising in the property expansions of the 
superfields containing fermions and bosons:
\begin{equation}
\Psi(x,\zeta,\bar{\zeta})\!=\!{\sum\sum}_{r\!+\!s~odd}(\zeta)^r(\bar{\zeta})^s
 \psi_{rs}(x),\,
\Phi(x,\zeta,\bar{\zeta})\!=\!{\sum\sum}_{r\!+\!s~even}(\zeta)^r(\bar{\zeta})^s
 \phi_{rs}(x).
\end{equation}
These components, labelled by $rs$, can be arranged in a 4$\times$4 magic 
square  where entries (as colour representations) tie in with the powers of 
chromicity. Thus we get the fermion  (see Table~\ref{ta1}) 
and boson (see Table~\ref{ta2}) squares below:

\begin{table}[th]
\tbl{Fermionic colour states $(\bar{\zeta})^r(\zeta)^s$; $r+s$ odd}
{\begin{tabular}{@{}|c|c|c|c|c|@{}}  
\hline
$r\backslash s$ & 0 & 1 & 2 & 3 \\ 
\hline
   0 &  & 3 &  &  1\\  
\hline
  1 & $\bar{3}$ &  & $3\oplus\bar{6}$ &  \\ 
\hline
   2 &  & $\bar{3}\oplus 6$ &   & 3\\ 
\hline
   3 &  1 &  & $\bar{3}$ &   \\
\hline
\end{tabular} \label{ta1}}
\end{table}  


\begin{table}[th]
\tbl{Bosonic colour states $(\bar{\zeta})^r(\zeta)^s$; $r+s$ even}
{\begin{tabular}{@{}|c|c|c|c|c|@{}} 
\hline
$r\backslash s$ & 0 & 1 & 2 & 3 \\ 
\hline
   0 &   1  &  &  $\bar{3}$ &   \\ \hline
   1 &    & $1\oplus 8$ &  & $\bar{3}$ \\ \hline
   2 & $3$ &   & $1\oplus 8$ &   \\ \hline
   3 &   & $3$ &   & 1  \\ 
\hline
\end{tabular}\label{ta2}}
\end{table}

\noindent Observe that reflection about the main diagonal is connected
with charge conjugation, whereas reflection about the cross-diagonal 
$(r,s) \leftrightarrow (3-s,3-r)$ does not affect the colour representation.

\subsection{Chromoduality and Colour Superfields}
We may take advantage of the last observation to cut down
the number of independent components arising in the property
expansion. For a start let the duality$^\times$  operation 
correspond to reflection about the cross-diagonal of the
monomials: $(\zeta)^r(\bar{\zeta})^s\rightarrow_\times
(\bar{\zeta})^{3-r}(\zeta)^{3-s}$. Specifically, using the
colour Levi-Civita symbol, we construct for example,
$(\zeta^{\bar\rho}\zeta^{\bar\sigma}\zeta^\lambda)^\times = 
\epsilon^{\bar{\rho}\bar{\sigma}\bar{\tau}}\zeta^\tau\,
\epsilon^{\lambda\mu\nu}\zeta^{\bar\mu}\zeta^{\bar\nu}/2$. 
Thus $(\zeta^1\zeta^2\zeta^3)^\times=
\zeta^1\zeta^2\zeta^3$  is colour selfdual, while
$1^\times = (\zeta^{\bar\sigma}\zeta^\sigma)^3/3!$ and vice versa.

So if we were to impose anti-selfduality on the magic square 
we would remain with two colour triplets (\& conjugates) in the 
fermion sectors (see Table~\ref{ta3}), with the colour sextet disappearing; 
the boson sector (see Table~\ref{ta4}) would comprise two singlets, 
one octet and one triplet (+ conjugate):
\begin{table}[th]
\tbl{Anti-selfdual $\psi$ with $r+s$ odd}
{\begin{tabular}{@{}|c|c|c|c|c|@{}}  \hline
$r\backslash s$ & 0 & 1 & 2 & 3 \\ \hline
      0 &   & 3 &  &   \\  \hline
     1 & $\bar{3}$ &  & $3'$ &  \\ \hline
      2 &  & $\bar{3'}$ &  & -\\ \hline
      3 &  &   & -  & \\  \hline
\end{tabular} \label{ta3}}
\end{table}  
\begin{table}[th]
\tbl{Anti-selfdual $\phi$ with $r+s$ even}
{\begin{tabular}{@{}|c|c|c|c|c|@{}}  \hline
$r\backslash s$ & 0 & 1 & 2 & 3 \\ \hline
   0 &  1  &  &  $\bar{3}$ &   \\  \hline
   1 &    & $1'\oplus 8'$ &  & -  \\ \hline
    2 &  3 &   &  -  &   \\ \hline
     3 &    & - &   & -  \\ \hline
\end{tabular} \label{ta4}}
\end{table}

\noindent The corresponding anti-selfdual (normalized) chromic expansions read, 
in detail ($Z\equiv \zeta^{\bar{\sigma}}\zeta^\sigma$)\,:
\begin{equation}
2\Psi(x,\zeta,\bar{\zeta}) =\left( \bar{\zeta}\psi(x)+\psi^c(x)\zeta\right) (1\!-\! Z^2/2)\! 
 +\!\left(\bar{\zeta}\psi'(x)+\psi'^c(x)\zeta\right)Z,
\end{equation}
\begin{equation}
2\overline{\Psi}(x,\zeta,\bar{\zeta}) \equiv \left(-\bar{\psi}(x)\zeta
 +\bar{\zeta}\overline{\psi^c}(x)\right)
(1\!-\! Z^2/2)\! 
 +\left(\bar{\psi'}(x)\zeta-\bar{\zeta}\,\overline{\psi'^c}(x) \right)Z,
\end{equation}
\begin{eqnarray}
2\Phi(x,\zeta,\bar{\zeta})&=&(\varphi(x)/\sqrt{2})(1-Z^3/6) +
  (\varphi' Z/\sqrt{6}+\zeta^{\bar{\sigma}}\phi'^{\sigma\bar{\rho}}\zeta^\rho/\sqrt{2})(1-Z) \nonumber \\
  & & +\;(\zeta^{\bar{\rho}}\zeta^{\bar{\sigma}}\phi^{\bar{\tau}}\epsilon^{\tau\sigma\rho}
  -\epsilon^{\bar{\rho}\bar{\sigma}\bar{\tau}}\phi^\tau\zeta^\sigma\zeta^\rho) (1-Z)/4.
\end{eqnarray}
Note also that both $\Psi$ and $\Phi$ are overall Bose but $\Psi$ carries spinorial 
indices $\alpha$ and that both left and right fermion fields occur in the 
magic square; so we are dealing with Dirac fields.  The salient points to stress after all 
these manoeuvres is that two fermion generations (each with three colours of fermions)
survive the antiduality constraint, while there are two scalar colour singlets $\varphi,\varphi'$.

\subsection{Gluon interactions}
The aim here is to show that the interactions with the gluons arise automatically 
from the supermetric $G$ or
frame vectors $E$ in the spacetime-chromicity sector (as listed in Appendix
A.3) through the full free Lagrangian
${\cal L} ={\cal L}_\Phi +{\cal L}_\Psi$, where
\begin{equation}
{\cal L}_\Phi=\frac{1}{2}\int\!d^3\zeta d^3\bar{\zeta} \sqrt{G..}
[G^{NM}\partial_M\Phi\partial_N\Phi + \mu^2\Phi^2]\quad{\rm and}
\end{equation}
\begin{equation}
{\cal L}_\Psi=\int\!d^3\zeta d^3\bar{\zeta}\, ({\rm sdet}{\cal E})
\bar{\Psi}[i\Gamma^A {E_A}^M\partial_M - m]\Psi.
\end{equation}

To simplify the discussion we shall at first ignore the property and spatial curvatures,
thereby setting all $c_i=0$, but will retain the gauge fields components of the metric
to check that the correct gluon interactions ensue. (Later on we will relax this
restriction as it can potentially complicate the picture by producing intergenerational 
mixing and finite wave function renormalization effects.) The normalization factors have
been correctly introduced in (10)-(12) in as much as
\[
\int\!d^3\zeta d^3\bar{\zeta}\,(\overline{\Psi}\Psi)=(\bar{\psi}\psi+\bar{\psi}'\psi')\;\;
{\rm and}\; \int\!d^3\zeta d^3\bar{\zeta}\,\,\Phi^2 = (\varphi^2+\varphi'^2
+\phi'^{\sigma\bar{\rho}}\phi'^{\rho\bar{\sigma}}+\phi^{\bar{\rho}}\phi^\rho).
 \]

The Bose field integration gives for the kinetic term
\begin{eqnarray}
2{\cal L}_{K\Phi}&=&\partial^m\varphi\partial_m\varphi+\partial^m\varphi'\partial_m\varphi'
+(\partial^m\bar{\phi}+ig_w\bar{\phi}W^m)(\partial_m\bar{\phi}-ig_wW_m\bar{\phi})\nonumber\\
 & & +{\rm Tr}\left[(\partial^m\phi'-ig_w[W^m,\phi'])(\partial_m\phi'-ig_w[W_m,\phi'])\right]
\end{eqnarray}
while the Fermi field produces
\begin{equation}
{\cal L}_{K\Psi}=i\bar{\psi}\gamma^m(\partial_m-ig_wW_m)\psi +
   i\bar{\psi}'\gamma^m(\partial_m-ig_wW_m)\psi',
\end{equation}
provided that $\Psi$ is a singlet in the auxiliary property gamma-matrix space,
as explained in ref. 2; this ensures that the only relevant derivative part of (13)
is $\gamma^a{E_a}^M\partial_M$, with $\Gamma^\alpha{E_\alpha}^M\partial_M$ 
giving zero. That is all well and good.
However an interesting byproduct of the analysis is that the property derivative 
contributions, 2$G^{\mu\bar{\nu}}\partial_{\bar{\nu}}\Phi\,\partial_\mu\Phi$, will
add to the masses by Planckian magnitudes
\[
(4/l^2)[-\varphi'^2+2\sqrt{3}\varphi\varphi'+\bar{\phi}\phi+{\rm Tr}(\phi'\phi')].
\]
At this stage we are unclear if this is a desirable feature or not, but we need to be 
reminded that there is also an unrenormalized  bare mass term $\mu^2\Phi^2$ 
which can serve to cancel out the Planckian additions; besides we have
disregarded the self-interactions such as $\Phi^4$ and $g\bar{\Psi}\Phi\Psi$.

\subsection{Gravitational interactions}
It is quite straightforward to introduce the space-time curvature via the full 
frame vectors (not just the gauge field part) as per eq (A.2):
\begin{equation}
\frac{1}{\sqrt{C}}\left(
\begin{array}{ccc}
{E_a}^m={e_a}^m  & {E_a}^\mu = ig_w (W_a\zeta)^\mu & 
\,\,(E_a)^{\bar{\mu}}\!=\!-ig_w(\bar{\zeta}W_a)^{\bar{\mu}}\\ 
{E_\alpha}^m=0 & {E_\alpha}^\mu ={\delta_\alpha}^\mu & {E_\alpha}^{\bar{\mu}}=0 \\ 
{E_{\bar{\alpha}}}^m=0 & {E_{\bar{\alpha}}}^\mu=0  & 
{E_{\bar{\alpha}}}^{\bar{\mu}}={\delta_{\bar{\alpha}}}^{\bar{\mu}}
\end{array}
\right)
\end{equation}
 or the full metric (6). Including the property curvature coefficients $c_i$,
 the Berezinian ($Z\equiv\bar{\zeta}\zeta$) reads:
 \begin{equation}
 {\rm sdet}{\cal E} =8(\sqrt{g..}/l^6)C^{-1}=8(\sqrt{g..}/l^6)[1-c_1Z+({c_1}^2\!-c_2)Z^2
                 +(2c_1c_2\!-\!c_3\!-\!{c_1}^3)Z^3].
 \end{equation}
Finally one must integrate over the property coordinates. 
The result of including space-time curvature is to produce a fully coordinconfident
ate
independent and gauge invariant Lagrangian unifying gravity and QCD. 
The only complication which arises 
is that the component fields, $\psi$ and $\phi$, may undergo wave function 
renormalizations determined by the property curvature coefficients $c_i$. For instance, 
noting that the spatial derivatives and the gauge field interactions are equally affected, the
fermion field $\psi$ must be rescaled by the factor $[1+ 3(5{c_1}^2-4c_2)/8]$, while the
other fermion colour triplet $\psi'$ does not require any rescaling. (Similar remarks 
apply to some of the boson field components.)

\section{\label{sec5} Discussion }
We see that it is a fairly simple matter to extend our formalism from our
previous papers to the strong force. Further, by including an overall gauge invariant 
factor $C$ multiplying our metric, we can generate a cosmological constant. Then 
considering a local SU(3) gauge transformation of our property coordinates
we end up producing an SU(3) Yang-Mills term in our Lagrangian. Together 
this unifies gravity with the strong force, as well as allowing for an unrestricted 
cosmological constant. As well, it automatically generates the correct interactions
of matter fields with gluons.

The next step in our program is to consider larger gauge groups, for instance 
U(1)$\times$SU(3) for dealing with electromagnetism plus the strong force, followed
by considerations of chirality to deal with the weak force. The restriction 
of the free parameters $c_i$ is also worth further study; in this paper we 
have confined ourselves to an overall factor $C$ multiplying the metric but 
there are likely to be other ways of limiting the extra degrees of freedom in the metric. 

We have been encountering issues with the time taken for algebraic computations, 
as the computational complexity is factorial in the number of property coordinates. 
This can be addressed by developing better algorithms to simplify expressions 
involving the metric
or through further analysis of the underlying mathematical structure to allow us to 
compute the Lagrangian in a simpler manner. As we are now confident our 
scheme is self consistent and frame invariant, we can take shortcuts in our 
calculations to determine the cosmological and gravitational constants. 
The ultimate goal of all this is to eventually unify gravity with the full standard 
model group of U(1)$\times$ SU(2)${}_L\times$SU(3), with enough restrictions 
on our free parameters $c_i$ to be able to predict the value of the cosmological 
constant, or at least to be consistent with observations while maintaining a 
minimal number of free parameters.

\appendix
\section{Trace identities}
In the evaluation of the Lagrangian we come across expressions of the form:
\[
\int (\bar\zeta X\zeta)(\bar\zeta Y \zeta)(\bar\zeta Z \zeta) d^3\zeta d^3{\bar\zeta}
\]
which will produce various traces over the matrices $X$, $Y$ and $Z$ when the 
property integration is performed. We now set about determining how to evaluate 
these expressions for $N$ property coordinates.

First, if we sum over $N$ property indices $\alpha_1, \alpha_2, ..., \alpha_N$, 
the following is true:
\[
\zeta^{\alpha_1} \zeta^{\alpha_2} ...  \zeta^{\alpha_N} = \varepsilon^{\alpha_1 \alpha_2
\ldots \alpha_N} \zeta^{1} \zeta^{2} ...\zeta^{N},
\]
where we take $\varepsilon^{12\ldots N} = 1$. This means that if we have some 
expression of the form:
\[
\zeta^{\bar\alpha_1} X_{1}{}^{\alpha_1 \bar\beta_1} \zeta^{\beta_1}  \zeta^{\bar\alpha_2} X_{2}{}^{\alpha_2 \bar\beta_2} \zeta^{\beta_2} ... \zeta^{\bar\alpha_{N}} X_{N}{}^{\alpha_{N} \bar\beta_{N}} \zeta^{\beta_{N}},
\]
we can permute this to:
\[
(\mathrm{sign}) X_{1}{}^{\alpha_1 \bar\beta_1}   
X_{2}{}^{\alpha_2 \bar\beta_2} ... X_{N}{}^{\alpha_{N} \bar\beta_{N}} 
\zeta^{\bar\alpha_1}\zeta^{\bar\alpha_2}...\zeta^{\bar\alpha_N}\zeta^{\beta_1} 
\zeta^{\beta_2}... \zeta^{\beta_N} .
\]
Where (sign) = $+1$ for $N = 1,4,5$ and $-1$ for $N = 2,3$ (+1 when 
$N$ mod 4 = 0 or 1, otherwise $-1$). We now rewrite this in terms of Levi-Civita symbols:
\[
(\mathrm{sign}) X_{1}{}^{\alpha_1 \bar\beta_1} X_{2}{}^{\alpha_2 
\bar\beta_2}\ldots X_{N}{}^{\alpha_{N} \bar\beta_{N}} \varepsilon^{\alpha_1 \alpha_2\ldots\alpha_N}
\varepsilon^{\bar\beta_1 \bar\beta_2\ldots\bar\beta_N} .
\]
This product of Levi-Civita symbols can be re-expressed as a matrix determinant, 
\begin{align*}
\varepsilon^{\alpha_1\alpha_2 \ldots \alpha_N}
\varepsilon^{\bar\beta_1\bar\beta_2\ldots\bar\beta_N} = 
\begin{array}{|cccc|}
\delta^{\alpha_1 \bar\beta_1} & \delta^{\alpha_1 \bar\beta_2}&\ldots&\delta^{\alpha_1 \bar\beta_N}\\
\delta^{\alpha_2 \bar\beta_1} & \delta^{\alpha_2 \bar\beta_2}&\ldots&\delta^{\alpha_2 \bar\beta_N}\\
\ldots& \ldots & \ldots & \ldots\\
\delta^{\alpha_N \bar\beta_1}&\delta^{\alpha_N \bar\beta_2}&\ldots& \delta^{\alpha_N\bar\beta_N}\\
\end{array}
\end{align*}
Thus we can find the results of integration over property coordinates for our 
non-abelian matrices $X_1,\ldots,X_N$. 

\subsection{SU(2) traces}
We now test this on SU(2), which we know already via explicit expansion to be:
\begin{align*}
\int \zeta^{\bar\mu} X^{\mu \bar\nu} \zeta^{\nu} \zeta^{\bar\alpha}
Y^{\alpha \bar\beta} \zeta^{\beta} d^2\zeta d^2\bar{\zeta}  = 
{\rm {\rm {\rm Tr}}}(XY) - {\rm {\rm {\rm Tr}}}(X) {\rm {\rm {\rm Tr}}}(Y)
\end{align*}
Using our above method, we have a minus sign since $N = 2$, and then we 
need the determinant:
\begin{align*}
\begin{array}{|cc|}
\delta^{\alpha_1 \bar\beta_1} & \delta^{\alpha_1 \bar\beta_2} \\
\delta^{\alpha_2 \bar\beta_1} & \delta^{\alpha_2 \bar\beta_2} \\
\end{array}
= \delta^{\alpha_1 \bar\beta_1} \delta^{\alpha_2 \bar\beta_2} - \delta^{\alpha_1 \bar\beta_2}\delta^{\alpha_2 \bar\beta_1} 
\end{align*}
Thus our expression becomes:
\begin{align*}
\zeta^{\bar\alpha_1} X^{\alpha_1 \bar\beta_1}  \zeta^{\beta_1} \zeta^{\bar\alpha_2}
Y^{\alpha_2 \bar\beta_2} \zeta^{\beta_2} =& - (\delta^{\alpha_1 \bar\beta_1} 
\delta^{\alpha_2 \bar\beta_2} -  \delta^{\alpha_1 \bar\beta_2}\delta^{\alpha_2 \bar\beta_1} )
 X^{\alpha_1 \bar\beta_1}Y^{\alpha_2 \bar\beta_2} \\
=& - {\rm {\rm Tr}}(X){\rm {\rm Tr}}(Y) + {\rm {\rm Tr}}(XY)
\end{align*}
after integration over property, which agrees with above.

\subsection{SU(3) traces}
We can now apply the same process to SU(3) traces. We will start in the 
permuted order, as this is what Mathematica deals with (i.e. no sign due to permutation). 
\begin{align*}
 X^{\alpha_1 \bar\beta_1} Y^{\alpha_2 \bar\beta_2} Z^{\alpha_3 \bar\beta_3} 
 \zeta^{\bar\alpha_1}  \zeta^{\bar\alpha_2} \zeta^{\bar\alpha_3} \zeta^{\beta_1} 
 \zeta^{\beta_2} \zeta^{\beta_3}  
\end{align*}
We need the determinant:
\begin{align*}
&\begin{array}{|ccc|}
\delta^{\alpha_1 \bar\beta_1} & \delta^{\alpha_1 \bar\beta_2}&\delta^{\alpha_1 \bar\beta_3} \\
\delta^{\alpha_2 \bar\beta_1} & \delta^{\alpha_2 \bar\beta_2}&\delta^{\alpha_2 \bar\beta_3} \\
\delta^{\alpha_3 \bar\beta_1} & \delta^{\alpha_3 \bar\beta_2}&\delta^{\alpha_3 \bar\beta_3} \\
\end{array}\\
=& \delta^{\alpha_1 \bar\beta_1} ( \delta^{\alpha_2 \bar\beta_2}\delta^{\alpha_3 \bar\beta_3} 
-  \delta^{\alpha_2 \bar\beta_3}\delta^{\alpha_3 \bar\beta_2} ) 
+ \delta^{\alpha_1 \bar\beta_2} (\delta^{\alpha_2 \bar\beta_3}\delta^{\alpha_3 \bar\beta_1} 
- \delta^{\alpha_2 \bar\beta_1} \delta^{\alpha_3 \bar\beta_3}) \\
&+ \delta^{\alpha_1 \bar\beta_3} ( \delta^{\alpha_2 \bar\beta_1} \delta^{\alpha_3 \bar\beta_2} 
-  \delta^{\alpha_2 \bar\beta_2}\delta^{\alpha_3 \bar\beta_1} )\\
=& \delta^{\alpha_1 \bar\beta_1}  \delta^{\alpha_2 \bar\beta_2}\delta^{\alpha_3 \bar\beta_3} 
-  \delta^{\alpha_1 \bar\beta_1}  \delta^{\alpha_2 \bar\beta_3}\delta^{\alpha_3 \bar\beta_2} 
 + \delta^{\alpha_1 \bar\beta_2} \delta^{\alpha_2 \bar\beta_3}\delta^{\alpha_3 \bar\beta_1} 
 - \delta^{\alpha_1 \bar\beta_2} \delta^{\alpha_2 \bar\beta_1} \delta^{\alpha_3 \bar\beta_3} \\
&+ \delta^{\alpha_1 \bar\beta_3}  \delta^{\alpha_2 \bar\beta_1} \delta^{\alpha_3 \bar\beta_2} 
-  \delta^{\alpha_1 \bar\beta_3} \delta^{\alpha_2 \bar\beta_2}\delta^{\alpha_3 \bar\beta_1} 
\end{align*} 
Thus after integration over property our expression becomes:
\begin{align*}
&{\rm Tr}(X) {\rm Tr}(Y) {\rm {\rm Tr}}(Z) - {\rm Tr}(X) {\rm Tr}(Y Z) + {\rm Tr}(X Y Z) \\
&- {\rm Tr}(X Y) {\rm Tr}(Z) + {\rm Tr}(X Z Y) - {\rm Tr}(X Z) {\rm Tr}(Y)
\end{align*}
For the purposes of Mathematica implementation there are several important 
subcases, namely when $X$, $Y$ or $Z$ are the identity. In this case their 
traces evaluate for $N= 3$:
for $Z= I$ we get ${\rm Tr}(X){\rm Tr}(Y) - {\rm Tr}(XY)$;
for $Y = Z= I$ we get $2 {\rm Tr}(X)$, and finally for $X=Y=Z=I$ we get 6. 

\subsection{Frame vectors and Vielbeins}
We have the overall curvature factor $C = 1 + c_1 \bar\zeta \zeta + c_2 (\bar\zeta \zeta)^2 + 
c_3 (\bar\zeta \zeta)^3$, 
leading to $\sqrt{C} = 1 + \frac{1}{2} c_1 \bar\zeta \zeta + \frac{1}{2} (c_2 - \frac{1}{4} c_1{}^2) 
(\bar\zeta\zeta)^2 + \frac{1}{2} ( c_3 - \frac{1}{2} c_1 c_2 + \frac{1}{8} c_1{}^3) (\bar\zeta\zeta)^3.$
In Section 4 we have taken upper-triangular frame vectors of the following 
form which produce the metric (6) used previously:
\begin{equation}
\mathcal{E}_M{}^{A} =
\sqrt{C} \left( \begin{array}{ccc}
 e_m{}^a & - i g_w W_m{}^{ \alpha \bar\nu} \zeta^{\nu} &  i g_w \zeta^{\bar\nu} W_m{}^{ \nu \bar\alpha} \\
0 & \delta_\mu{}^\alpha &  0\\
0 & 0 &\delta_{\bar\mu}{}^{\bar\alpha} \\
\end{array}
\right).
\end{equation}
These have as their inverse the vielbein matrix,
\begin{equation}
E_A{}^{M} =
\frac{1}{\sqrt{C}}
\left(
\begin{array}{ccc}
 e_a{}^m &  i g_w  W_a{}^{ \mu \bar\nu} \zeta^{\nu} & - i g_w \zeta^{\bar\nu} W_a{}^{ \nu \bar\mu} \\
0 & \delta_\alpha{}^\mu &  0\\
0 & 0 &\delta_{\bar\alpha}{}^{\bar\mu} 
\end{array}
\right).
\end{equation}

\end{document}